\documentclass[12pt]{article}
\usepackage{amsmath}
\usepackage{graphicx}
\usepackage{floatflt}
\usepackage{subfigure}
\usepackage{theorem}
\usepackage{url}

\newcommand{\tr}{\mathrm{tr}}
\newcommand{\bra}[1]{\langle#1|}
\newcommand{\ket}[1]{|#1\rangle}
\newcommand{\braket}[2]{\langle#1|#2\rangle}
\newcommand{\ketbra}[2]{|#1\rangle\langle#2|}
\newcommand{\interm}[0]{|X|^2}

\newtheorem{lemma}{Lemma}
\newtheorem{theorem}{Theorem}
\newtheorem{example}{Example}
\newtheorem{proposition}{Proposition}
\newtheorem{definition}{Definition}
\newtheorem{corollary}{Corollary}

\sloppypar

\begin{document}

\title{Improving Ranking Using Quantum Probability}
\ifx\proof\undefined
\newcommand{\qed}{\hfill \mbox{\raggedright \rule{.07in}{.1in}}}
\newenvironment{proof}{\vspace{1ex}\noindent{\bf Proof}\hspace{0.5em}}
{\hfill\qed\vspace{1ex}\linebreak}
\newenvironment{pfof}[1]{\vspace{1ex}\noindent{\bf Proof of #1}\hspace{0.5em}}
{\hfill\qed\vspace{1ex}}
\author{Massimo Melucci\\
  {\normalsize University of Padua}\\
  {\normalsize\url{massimo.melucci@unipd.it}}}
\else
\numberofauthors{1}
\author{
  \alignauthor
  Massimo Melucci\\
  \affaddr{University of Padua}\\
  \email{massimo.melucci@unipd.it}
}
\conferenceinfo{PODS}{'11 Athens, Greece}
\fi

\maketitle
\newpage

\begin{abstract}
  Data management systems, like database, information extraction, information retrieval or
  learning systems, store, organize, index, retrieve and rank information units, such as
  tuples, objects, documents, items to match a pattern (e.g. classes and profiles) or meet a
  requirement (e.g., relevance, usefulness and utility).  To this end, these systems rank
  information units by probability to decide whether an information unit matches a pattern or
  meets a requirement.  Classical probability theory represents events as sets and probability
  as set measures. Thus, distributive and total probability laws are admitted. Quantum
  probability is a non-classical theory nor does admit distributive and total probability
  laws. Although ranking by probability is far from being perfect, it is optimal thanks to
  statistical decision theory and parameter tuning.

  The main question asked in the paper is whether further improvement over the optimality
  provided by probability may be obtained if the classical probability theory is replaced by
  quantum probability theory.  Whereas classical probability (and detection theory) is based
  on sets such that the regions of acceptance / rejection are set-based detectors, quantum
  probability is based on subspace-based detectors.

  The paper shows that ranking information units by quantum probability differs from ranking
  them by classical probability provided the same data used for parameter estimation.  As
  probability of detection (also known as recall or power) and probability of false alarm
  (also known as fallout or size) measure the quality of ranking, we point out and show that
  ranking by quantum probability yields higher probability of detection than ranking by
  classical probability provided a given probability of false alarm and the same parameter
  estimation data.

  As quantum probability provided more effective detectors than classical probability within
  other domains that data management, we conjencture that, the system that can implement
  subspace-based detectors shall be more effective than a system which implements a set-based
  detectors, the effectiveness being calculated as expected recall estimated over the
  probability of detection and expected fallout estimated over the probability of false alarm.
\end{abstract}
\newpage

\section{Introduction}
\label{sec:introduction}

Data management systems, like database, information retrieval (IR), information extraction
(IE) or learning systems, store, organize, index, retrieve and rank information units, like
tuples, objects, documents, items.  A wide range of applications of these systems have emerged
that require the management of uncertain or imprecise data.  Important examples of data are
sensor data, webpages, newswires, imprecise attribute values.  What is common to all these
applications is uncertainty and then that they have to deal with decision and statistical
inference.

Ranking is perhaps the most crucial task performed by the data management systems which have
to deal with uncertainty. In many applications, ranking aims at deciding or inferring, for
example, the class assigned to a unit or the order by relevance, usefulness, or utility of the
units delivered to another application or to an end user.  In addition, ranking is performed
to decide whether a unit is placed at a given rank.

The management of imprecise data require means for ranking information units by probability.
Ranking places information units in a list ordered by a measure of utility, cost, relevance,
etc..  A probability theory measures the uncertainty of the decision.  To this end, the
definition of an event space and the estimation of probabilities are necessary steps for
representing imprecise data and making predictions within many contexts of data management
like machine learning, information retrieval or probabilistic databases.  

The measurement of the imprecision and the uncertainty in the data leads to the definition of
regions of acceptance of a predefined set of hypotheses, thus bringing many decision problems
to the calculation of a probability of detection and of a probability of false alarm. Although
the data management systems reach good results thanks to classical probability theory and
parameter tuning, ranking is far from being perfect because useless units are often ranked at
the top or useful units are missed.

Classical probability theory describes events and probability distributions using sets and set
measures, respectively, according to Kolmogorov's axioms~\cite{Kolmogorov56}.  In contrast,
quantum probability theory describes events and probability distributions using Hermitian
operators in the complex Hilbert vector space. Whereas parameter tuning is performed within a
fixed probability theory, the adoption of quantum probability entails a radical change.
Furthermore, whereas classical probability is based on sets such that the regions of
acceptance or rejection are set-based detectors (i.e., indicator functions), quantum
probability is based on subspace-based detectors and the detectors are projector-based. Note
that the use of quantum probability does not imply that quantum phenomena are investigated in
the paper; we are interested in the formalism based the Hilbert vector spaces instead.

The main question asked in the paper is whether further improvement may be obtained if the
classical probability theory is replaced by the quantum probability theory. The paper shows
that ranking information units by quantum probability yields different outcomes which are in
principle more effective than ranking them by classical probability given the same data
available for parameter estimation.  The effectiveness is measured in terms of probability of
detection (also known as recall or power) and probability of false alarm (also known as
fallout or size).

We structure the paper as follows. Section~\ref{sec:class-prob-quant} illustrates the basics
of the probability theory through a view that encompasses both theories.
Section~\ref{sec:quant-prob-detect} compares quantum detection with classical
detection. Section~\ref{sec:optim-rank-quant} shows that the ranking by quantum probability
more effective than the ranking by classical probability. Section~\ref{sec:interpr-quant-proj}
provides an interpretation of the projectors which define the regions of acceptance and
rejection. Section~\ref{sec:impl-rank} describes the algorithm for ranking information units
by quantum probability. Section~\ref{sec:related-work} provides an overview of the related
work.

\begin{figure}[t]
  \centering
  \includegraphics[width=0.750\columnwidth]{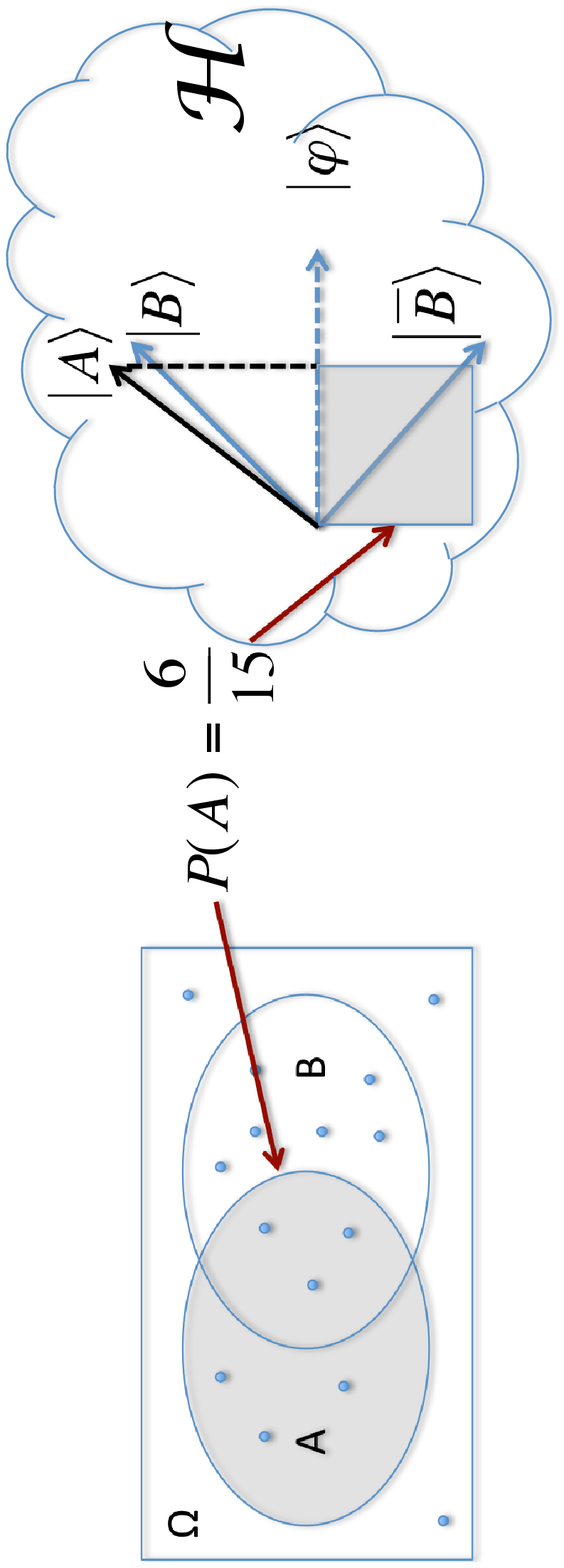}
  \caption{In the classical probabilistic theory, events (e.g., feature occurrences,
    category memberships, review scores, location, task, genre) are represented as sets
    and the probability measure is based on a set measure, e.g., set cardinality. In
    contrast, in quantum probability, events are represented as orthonormal vectors and
    the probability measure is the trace of the product between a density matrix and the
    matrix representing an event.  The simple example in the figure depicts that when
    vectors are used to implement both events and densities the probability in the vector
    space is the squared size of the projection of $\ket{A}$ onto $\ket{\varphi}$.}
  \label{fig:correspondence}
\end{figure}

\section{Classical Probability and Quantum Probability}
\label{sec:class-prob-quant}

In this section, we introduce a special view of probability distributions for the classical
theory of probability. The same view is also introduced for quantum probability, which is a
non-classical theory and does not admit the distributive law, to provide a general framework
for quantum and classical probabilities; the view is depicted in
Figure~\ref{fig:correspondence}.

Before introducing the view of probability theory, some basic definitions are provided.  A
probability space is a set of mutually exclusive events such that each event is assigned a
probability between $0$ and $1$ and the sum of the probabilities over the set of events is
$1$. For the sake of clarity, we introduce the case of binary event spaces because
it is the simplest and most common in data management~--~keyword occurrence in webpages,
binary features in sample records or binary attribute values in relational tables are some
examples.  The case of binary event spaces are usually represented by mutually exclusive
scalars like $0$ and $1$. If binary scalars are used, the mutual exclusiveness is given
by the scalar product, for example, $01 = 0$ (see~\cite{Boole1854}).

Whereas the scalars $\{0,1\}$ is a possible representation of events, the vectors of a complex
finite-dimensional are another option.  When using vectors, an event is $\left(0, 1\right)'$
and its complement is $\left(1, 0\right)'$.  The representation of the events must encode the
mutual exclusiveness.  If binary vectors are used, the mutual exclusiveness is given by the
inner product, i.e., $\left(0, 1\right) \left(1, 0\right)' = 0$.

When the event space is not binary (e.g., when the events are represented by $k$ natural
numbers $0, 1, \ldots, k-1$), a binary representation can again be used.  The vector $\left(0,
  \cdots, 0, 1\right)'$ is assigned to symbol $0$, the vector $\left(0, \cdots, 1, 0\right)'$
is assigned to symbol $1$, and so on until $\left(1, \cdots, 0, 0\right)'$ is assigned to
symbol $k-1$ .  Whatever the representation is used, the inner product between two vectors
must be $0$ and their norm must be $1$.

The mapping between the probabilities and the events is called ``probability distribution''
which is a function mapping a mathematical object which represents an event to a real number
ranging between $0$ and $1$.  The difference between classical probability and quantum
probability; the difference is due to the way the event space and the probability distribution
are represented.

The starting point of the view of probability used in the paper is the algebraic form of the
probability space.  To this end, Hermitian\footnote{``Symmetric'' is adopted in the real
  field.} (or self-adjoint) linear operators are used.  In quantum mechanics, ``operator'' is
preferred to ``matrix'' yet in the paper, for the sake of clarity, ``matrix'' is preferred
because, for a fixed basis, the matrices are isomorphic to the operators. A matrix is
Hermitian when it is equal to its conjugate transpose. Hermitian matrices are important
because their eigenvalues are always real.  In particular, Hermitian matrices with trace $1$
is the key notion in quantum probability because the sum of the eigenvalues is $1$ and, thus,
the eigenvalues can be viewed as a probability distribution.  

The projector is an idempotent Hermitian matrix. Every subspace has one projector and then the
projectors are 1:1 correspondence with the subspaces. Each vector corresponds to one projector
with rank one defined as the outer product of the vector by its conjugate transpose.  There
are two main instructions for representing events using projectors:
\begin{itemize}
\item the projectors must be mutually orthogonal for representing the mutual exclusiveness of
  the events, and
\item the projectors must have trace $1$ for making probability calculation consistent with
  the probability axioms.
\end{itemize}

An event space and a probability function defined over it are represented using Hermitian
matrices with trace $1$.  In particular, a projector represents an event and an event space is
modeled by a collection of projectors.  As the union of the events results in the whole event
space, the sum of the projectors of a collection corresponding to an event space results in
the unity.  More specifically, if $\left\{\mathbf{E}_0, \ldots, \mathbf{E}_{k-1}\right\}$ is a
collection of mutually orthogonal projectors,
\begin{equation*}
  \mathbf{E}_0 + \ldots + \mathbf{E}_{k-1} = \mathbf{I}\ ,
\end{equation*}
the latter being termed ``resolution to the unity''.  
For example, using the Dirac notation introduced in Appendix~\ref{sec:dirac-notation}, the
projector of two events are represented by
\begin{equation}
  \label{eq:22}
 \ket{1} =
  \left(
    \begin{array}{c}
      1 \\ [5pt] 0
    \end{array}
  \right)
  \qquad
  \ket{0} =
  \left(
    \begin{array}{c}
      0 \\ [5pt] 1
    \end{array}
  \right)
\end{equation}
and
\begin{equation}
  \label{eq:26}
  \ketbra{1}{1} =
  \left(
    \begin{array}{cc}
      1 & 0 \\ [5pt]
      0 & 0
    \end{array}
  \right)
  \qquad
  \ketbra{0}{0} =
  \left(
    \begin{array}{cc}
      0 & 0 \\ [5pt]
      0 & 1
    \end{array}
  \right)
\end{equation}
However, there is not a unique representation of an event space.  For example, the following
vectors are also representing mutually exclusive events:
\begin{equation}
  \label{eq:27}
  \left(
    \begin{array}{c}
      \frac{1}{\sqrt{2}} \\ [5pt] 
      \frac{1}{\sqrt{2}} 
    \end{array}
  \right)
  \qquad
  \left(
    \begin{array}{r}
      \frac{1}{\sqrt{2}} \\ [5pt] 
      -\frac{1}{\sqrt{2}} 
    \end{array}
  \right)
\end{equation}
thus leading to a different resolution to the unity given by the following projectors
\begin{equation}
  \label{eq:28}
  \left(
    \begin{array}{cc}
      \frac{1}{2} & \frac{1}{2} \\ [5pt]
      \frac{1}{2} & \frac{1}{2}
    \end{array}
  \right)
  \qquad
  \left(
    \begin{array}{rr}
      \frac{1}{2} & -\frac{1}{2} \\ [5pt]
      -\frac{1}{2} & \frac{1}{2}
    \end{array}
  \right)  
\end{equation}
The second kind of Hermitian matrix of a probability space is the density matrix; the density
matrix encapsulates the probability values assigned to the events.  In Physics, a density
matrix represents the \emph{state} of a microscopic system, such as a particle, a photon,
etc..  The structure of a microscopic system is unknown. Yet a device can measure the system
to obtain some information. A microscopic system is similar to a urn of colored balls. The
internal composition of the urn is always unknown even if opened and observed because the
device disturbs the state (i.e., the distribution of the colors) of the urn.

In data management and in other domains different from Particle Physics, a system is
macroscopic instead. \footnote{A macroscopic system in data management is thus not a computer
  systems as far as it is concerned in the paper.}  Examples of macroscopic systems in data
management are webpages, customers, queries, clicks, tuples, attributes, and so on. The states
of these systems correspond to the probability densities according to which keywords, reviews,
attribute values are observables to be measured from such systems.
Density matrices are a powerful formalism in the macroscopic worlds too because they allow us
to introduce the algebraic approach adopted for incorporating the more powerful probability
space and decision rule suggested in the paper.

To the end of introducing the way density matrices are defined, consider two equiprobable
events, e.g., the occurrence of a feature or a positive/negative customer review.  The
probability distribution is $\left(\frac{1}{2}, \frac{1}{2}\right)$ where each value refers to
an event.  As an alternative to a list, the probability distribution can be arranged along the
diagonal of a two-dimensional matrix and the other matrix elements are zeros.  For example,
the matrix corresponding to the probability distribution of two equally probable events is
\begin{equation*}
  \mu =
  \left(
    \begin{array}{cc}
      \frac{1}{2} & 0 \\ [5pt]
      0 & \frac{1}{2}
    \end{array}
  \right) 
\end{equation*}
In general, the probability distribution $\left(p_1, \ldots, p_k\right)$ of a $k$-event space
can be written as
\begin{equation*}
  \mu =
  \left(
    \begin{array}{cccc}
      p_1 	& 0 	& \cdots 	& 0 	\\ [5pt]
      0   	& p_2	& \cdots 	& 0 	\\ [5pt]
      \vdots   	& \vdots& \ddots	& \vdots\\ [5pt]
      0  	& 0	& \cdots 	& p_k 	\\ [5pt]
  \end{array}
  \right) 
\end{equation*}
A probability distribution is \emph{pure} when the density matrix is a projector, otherwise,
the distribution is \emph{mixed}.  A distribution is mixed when the density matrix is a
mixture of density matrices; a pure distribution is an instance of mixture with one matrix.
The density matrix representing a pure distribution is 1:1 correspondence with a density
vector such that the projector is the outer product between the vector and its conjugate
transpose. A classical probability distribution is pure when the probability is concentrated
on a single elementary event which is the certain event and then has probability $1$.

Given a density matrix, the spectral theorem helps find the underlying events and the related
probabilities. Because of the importance of the spectral theorem, we provide its definition
below:
\begin{theorem}
  \label{sec:class-prob-quant-1}
  To every Hermitian matrix $\mathbf{A}$ on a finite-dimensional complex inner product
  space there correspond real numbers $\alpha_0, \ldots, \alpha_{r-1}$ and rank-one projectors
  $\mathbf{E}_0, \ldots, \mathbf{E}_{r-1}$ so that the $\alpha_j$'s are pairwise distinct, the
  $\mathbf{E}_j$ are mutually orthogonal, $\sum_{j=0}^{r-1} \mathbf{E}_j = \mathbf{I}$,
  $\sum_{j=0}^{r-1} \alpha_j = 1$ and $\sum_{j=0}^{r-1} \alpha_j \mathbf{E}_j = \mathbf{A}$.
\end{theorem}
\begin{proof}
  See~\cite[page 156]{Halmos87}.
\end{proof}
The eigenvalues are the spectrum and $\mathbf{E}_0, \ldots, \mathbf{E}_{r-1}$ are the
projectors of the spectrum of $\mathbf{A}$. From Theorem~\ref{sec:class-prob-quant-1}, thus, a
pure distribution is always a rank-one projector.

The spectral theorem says that any Hermitian matrix corresponding to a distribution can be
decomposed as a linear combination of projectors (i.e. pure distributions) where the
eigenvalues are the probability values associated to the events represented by the
projectors. The eigenvalues are real because the decomposed matrix is Hermitian, are
non-negative and sum to $1$. For example, when the matrix corresponding to the distribution of
two equally probable events is considered, the spectral theorem says that
\begin{equation*}
\mu =
  \left(
    \begin{array}{cc}
      \frac{1}{2} & 0 \\ [5pt]
      0 & \frac{1}{2}
   \end{array}
  \right)
  =
  \frac{1}{2}
  \left(
    \begin{array}{cc}
      1 & 0 \\ [5pt]
      0 & 0
   \end{array}
  \right)
  +
  \frac{1}{2}
  \left(
    \begin{array}{cc}
      0 & 0 \\ [5pt]
      0 & 1
   \end{array}
  \right)
\end{equation*}
A mixed distribution have more non-zero eigenvalues, a pure distribution has a single
eigenvalue $1$.


In classical probability, every pure distribution represented by a diagonal density matrix
corresponds to a projector. However, in general, a density matrix is not necessarily diagonal,
yet the matrix is necessarily Hermitian. For example,~\eqref{eq:28} are trace-one projectors
and correspond to pure distributions, thus there is a certain event (with probability $1$) and
an impossible event (with probability $0$, of course).  Yet, they are not diagonal. When, for
example, keyword occurrence in webpages is represented, the first projector may be assigned
eigenvalue $1$ and the other is assigned eigenvalue $0$. Thus the former represents the
certain event and the latter represents the impossible event in the probability space.

When using the algebraic form to represent probability spaces, the function for computing a
probability is the trace of the matrix obtained by multiplying the density matrix by the
projector corresponding to the event. The usual notation for the probability of the event
represented by projector $\mathbf{E}$ when the distribution is represented by density matrix
$\rho$ is
\begin{equation}
  \label{eq:1}
  \tr(\rho \mathbf{E})
\end{equation}
also known as Born's rule~\cite{vonNeumann55}.  For example, when $\rho = \mu$
\begin{equation*}
\mu = 
\left(
  \begin{array}{cc}
    \frac{1}{2} & 0 \\ [5pt]
    0 & \frac{1}{2}
  \end{array}
\right)
\qquad
\mathbf{E} = 
\left(
  \begin{array}{cc}
    1 & 0 \\ [5pt]
    0 & 0
  \end{array}
\right)
\end{equation*}
the probability is
\begin{equation*}
\tr(\mu \mathbf{E}) = \tr
\left(
  \left(
    \begin{array}{cc}
      \frac{1}{2} & 0 \\ [5pt]
      0 & \frac{1}{2}
    \end{array}
  \right)
  \left(
    \begin{array}{cc}
      1 & 0 \\ [5pt]
      0 & 0
    \end{array}
  \right)
\right) = \tr
\left(
  \left(
    \begin{array}{cc}
      \frac{1}{2} & 0 \\ [5pt]
      0 & 0
    \end{array}
  \right)
\right) = \frac{1}{2}
\end{equation*}
When $\mathbf{E} = \ketbra{x}{x}$ is a rank-one projector, the trace-based probability
function can be written as
\begin{equation*}
\tr(\rho \mathbf{E}) = \bra{x}\rho\ket{x}
\end{equation*}
When $\rho$ is a rank-one projector $\ketbra{y}{y}$, then
\begin{equation*}
\tr(\rho \mathbf{E}) = |\braket{x}{y}|^2
\end{equation*}
From the example, the definition of a function that computes the probability of an event when
the probability is already allocated in the diagonal of the density matrix may be
odd. However, we have shown that not all the density matrices corresponding to a distribution
need to be diagonal matrices and the diagonal elements do not necessarily correspond to
probability values, although they do have to sum to $1$.

A density matrix encapsulate the values assigned to the events by a probability function
because of Gleason's theorem stated below and proved in~\cite{Hughes89}.
\begin{theorem}
  To every probability distribution on the set of all projectors in a complex vector space
  with dimension greater than $2$ there corresponds a unique density matrix $\rho$ on the same
  vector space for which the probability of the event represented by a projector
  $\ketbra{x}{x}$ is $\tr(\rho \ketbra{x}{x})$ for every unit vector $x$ in the vector space.
\end{theorem}
Basically, the theorem tells us that corresponding to a probability distribution is one
density matrix such that the probability of any event represented as a projector is
calculated by the trace function.


The probability of an event when computed using a mixture differs from the probability
computed using a pure state, yet they share the classical probability term whereas the
difference is called \emph{interference term}.  Using a mixture,
\begin{equation}
  \label{eq:43}
  \tr(\mu \ketbra{x}{x}) = |a_0|^2|b_0|^2 + |a_1|^2|b_1|^2 \qquad \mu = 
  \left(
    \begin{array}{cc}
      |a_1|^2 & 0  \\ [6pt]
      0 & |a_0|^2   
    \end{array}
  \right)
\end{equation}
Using superposition,
\begin{equation}
  \label{eq:8}
  \tr(\rho \ketbra{x}{x}) = |a_0|^2|b_0|^2 + |a_1|^2|b_1|^2 + 2|a_0||b_0||b_1||b_0|\cos\theta
\end{equation}
where $\rho = \ketbra{v}{v}$ and $\theta$ is the angle of the polar representation of the
complex number $a_0 \bar{b}_0 a_1 \bar{b}_1$.  Suppose, as an example, that $\ket{x}$
represents the event ``the keyword occurs'' and the density matrix represents the probability
distribution of keyword occurrence in useful webpages.  The common factor (i.e.~\eqref{eq:43})
is the sum of two probabilities; the probability that the webpage is not useful ($|a_0|^2$)
multiplied by the probability that the keyword occurs in a useless webpage ($|b_0|^2$), and
the probability that the webpage is useful ($|a_1|^2$) multiplied by the probability that the
keyword occurs in a useful webpage ($|b_1|^2$). The sum is nothing but an application of the
law of total probability.

The quantity $2|a_0||b_0||b_1||b_0|\cos\theta$ is the interference term. As the interference
term ranges between $-1$ and $+1$, the probability of keyword occurrence computed when
usefulness is superposed with uselessness becomes different from the common factor in which
usefulness and uselessness are mutually exclusive and their probability distribution is
described by a mixture. The interference term can be so large that the law of total
probability is violated and any probability space obeying Kolmogorov's axioms cannot admit the
probability values $|a_1|^2$ and $|b_1|^2$, thus requiring the adoption of a quantum
probability space~\cite{Accardi97,Accardi&82}.


\section{Quantum Probability and Decision}
\label{sec:quant-prob-detect}

In general, the information stored in the data is acquired and delivered through
information unit representation and ranking, these processes are described in terms of
decision and estimation, and they are therefore affected by error. The error could be
eliminated only if precise and exhaustive methodological tools and computer systems were
developed.  Nevertheless, there is a trade-off between precision, exhaustivity and the
computation cost because high level of the former can be achieved only if a high
computation cost is devoted.  Thus, a certain amount of error is unavoidable yet can be
controlled and limited below a given threshold.

Either a set of statements, or hypotheses, must be decided to best describe the
information unit insofar as data permit to judge (e.g., the best topic(s) to which a webpage is
assigned), or the values of certain quantities (also known as parameters)
characterizing the information unit must be estimated, the probability of detection and
the probability of false alarm related to a decision must be calculated. In the paper, a
great deal of attention is paid to decision whereas estimation is set apart not because
estimation is little important, but because estimation would require another research
stream had it to be addressed to the appropriate level of exhaustivity.

Many tasks in data management are decision problems, examples are the classification of
images with respect to predefined patterns, the categorization of webpages to topics,
contextual advertising (i.e., the decision whether an ad has to displayed in a
search engine result page), the retrieval and ranking of webpages (i.e., the decision as
to whether a webpage has to put at rank $r$ of a search engine result page), probabilistic
databases (i.e., the decision about the correct value of an attribute and then the
computation of the associated probability).

Our illustration of decision theory is necessarily brief and confined to its simplest
aspects and examples.  The illustration is also organized in such a way as to bring out
most clearly the parallels between classical probability-based decision and quantum
probability-based decision.  The examples are chosen from elementary information retrieval
or machine learning theory and perhaps provide a basis for comparison with the quantum
case.

A certain information unit (e.g., a webpage or an store item) is observed in such a way as
to obtain numbers (e.g., the PageRank or the number of positive reviews) on the basis of
which a decision has to be made about its state. The numbers observed are, for example,
the frequency of a feature in the information unit, the simplest example being the
frequency of a keyword in a webpage used for calculating search engine statistical ranking
functions.  For the sake of clarity, we use the binary frequency and the feature
presence/absence case in the paper. The state might be, for example, the relevance of the
webpage to the search engine user's interests or the customer's willingness to buy the
store item. The use of the term ``state'' is not coincidental because the numbers are
observed depending upon the density matrix, which is indeed the mathematical notion
implementing the state of a system. Thus, quantum probability ascribes the decision about
the state of an information unit to testing the hypothesis that the density matrix has
generated the observed numbers.

Consider the hypothesis that the state of the system is the density matrix $\rho_1$ and
the alternative hypothesis that the state of the system is the density matrix $\rho_0$.
The two hypotheses can be labeled $H_1$ and $H_0$, respectively.  In data management,
hypothesis $H_0$ asserts, for example, that a customer does not buy an item or that a
webpage shall be irrelevant to the search engine user whereas hypothesis $H_1$
asserts that an item shall be bought by a customer or that a webpage shall be relevant to
the user.  Therefore, the probability that, say, a feature occurs in an item which shall
not be bought by a customer or a keyword occurs in a webpage which shall be irrelevant to
the search engine user depends on the state (i.e., the density matrix).


Statistical decision theory is a old topic and Neyman-Pearson's lemma is by now one out of the
most important results which provides a criterion for deciding upon hypotheses instead of the
Bayesian approach.  The lemma provides the rule to govern the decider's behaviour and decide
the true hypothesis without hoping to know whether it is true. Given an information unit and
an hypothesis about the unit, such a rule calculates a specified number (e.g., a feature) and,
if the number is greater than a threshold reject the hypothesis, otherwise, accept it.  Such a
rule tells nothing whether, say, the item shall be bought by the customer, but the lemma
proves that, if the rule is followed, then, in the long run, the hypothesis shall be accepted
at the highest probability of detection (or power) possible when the probability of false
alarm (or size)~\cite{Neyman&33} is not higher than a threshold. The set of the pairs given by
size and power is the power curve which is also known as the Receiver Operating Characteristic
(ROC) curve.

Neyman-Pearson's lemma implies that the set of the observable numbers (e.g., features) can
be partitioned into two distinct regions; one region includes all the numbers for which
the hypothesis shall be accepted and is termed acceptance region, the other region includes
all the numbers for which the hypothesis shall be rejected and is termed rejection region.
For example, if a keyword is observed from webpages and only presence/absence is observed,
the set of the observable numbers is $\{0, 1\}$ and each region is one out of possible
subsets, i.e., $\emptyset, \{0\}, \{1\}, \{0,1\}$.  



The paper reformulates Neyman-Pearson's lemma in terms of subspaces instead of subsets to
utilize quantum probability.  Therefore, the region of acceptance and the region of
rejection must be defined in terms subspaces.  In the following, we illustrate the
algorithm for calculating the most efficient test in Hilbert spaces.  The following result
holds:
\begin{theorem}
  \label{the:helstrom}
  Let $\rho_1, \rho_0$ be the density matrices under $H_1, H_0$, respectively.  The region
  of acceptance at the highest power at every size is given by the projectors of the
  spectrum of
  \begin{equation}
    \label{eq:5}
    \rho_1 - \lambda\rho_0 \qquad \lambda > 0
  \end{equation}
  whose eigenvalues are positive.
\end{theorem}
\begin{proof}
  See~\cite{Helstrom76}.
\end{proof}
\begin{definition}
  An optimal projector is a projector which identifies the region of acceptance and the
  region of rejection according to Theorem~\ref{the:helstrom}.
\end{definition}
\begin{definition}
  We define the \emph{discriminant function} as
  \begin{equation}
    \label{eq:32}
    \tr((\rho_1 - \lambda\rho_0)\mathbf{E})
  \end{equation}
  where $\mathbf{E}$ is a projector. If the discriminant function is positive, the
  observed event represented by $\mathbf{E}$ is placed in the region of acceptance.
\end{definition}
Suppose that the density matrix that corresponds to $H_1$ is a mixed, classical
probability distribution. The mixed case is the usual method for dealing with uncertainty
in data management, even though more than one feature may exist or the feature may not be
binary; however, the number of features or the number of values of a feature is not
essential in the paper.  Let $\mu_1$ be such a mixed distribution and
\begin{equation}
  \label{eq:6}
  \mu_1 = p_1 \mathbf{P}_1 + (1-p_1)\mathbf{P}_0 \\
  {   } = 
  \left(
    \begin{array}{cc}
      p_1 & 0 \\ [5pt]
      0 & 1-p_1
    \end{array}
  \right)
\end{equation}
where
\begin{equation}
  \label{eq:11}
  \mathbf{P}_1 = 
  \left(
    \begin{array}{cc}
      1 & 0 \\ [5pt]
      0 & 0
    \end{array}
  \right)
  \qquad 
  \mathbf{P}_0 = 
  \left(
    \begin{array}{cc}
      0 & 0 \\ [5pt]
      0 & 1
    \end{array}
  \right)
\end{equation}
Similarly,
\begin{equation}
  \label{eq:7}
  \mu_0 = p_0 \mathbf{P}_1 + (1-p_0)\mathbf{P}_0 \\
  {   } = 
  \left(
    \begin{array}{cc}
      p_0 & 0 \\ [5pt]
      0 & 1-p_0
    \end{array}
  \right)  
\end{equation}
When $\mathbf{P}_1$ is observed, the power and the size are, respectively,
\begin{equation}
  \label{eq:21}
  P_d = \tr(\mu_1\mathbf{P}_1) = p_1 \qquad P_0 = \tr(\mu_0\mathbf{P}_1) = p_0
\end{equation}
In the classical case, $\mathbf{P}_0, \mathbf{P}_1$ represents the absence and the
presence, respectively, of a feature.  Hence, the possible acceptance or rejection regions
are $\mathbf{0}, \mathbf{P}_0, \mathbf{P}_1, \mathbf{I}$ which correspond respectively to
``never accept'', ``accept when the feature does not occur'', ``accept when the feature
occurs'' and ``always accept''.  Thus, the decision on, say, webpage classification, topic
categorization, item suggestion, can be made upon the occurrence of one or more features
because $\mathbf{P}_0, \mathbf{P}_1$ represent ``physical'' events. Furthermore, the
discriminant function in the mixed case is
\begin{equation}
  \label{eq:13}
  \tr((\mu_1 - \lambda\mu_0)\mathbf{E}) \qquad \mathbf{E} \in \left\{ \mathbf{0},
      \mathbf{P}_0, \mathbf{P}_1, \mathbf{I} \right\}
\end{equation}
The power curve can be built as follows.  Suppose, as an example, that a keyword describes
webpage content and that that webpage either includes ($\mathbf{P}_1 = \ketbra{1}{1}$) or
does not include ($\mathbf{P}_0 = \ketbra{0}{0}$) the keyword. $P_0 = 0$ only if $P_d = 0$
and this point corresponds to the event represented by $\mathbf{0}$.  Let $p_1, p_0$ be
the probability that the keyword occurs in a relevant webpage or in a non-relevant
webpages, respectively.  When the keyword is the unique observed feature, the webpage is
presented to the user if $p_1 > \lambda p_0$ and the keyword occurs, or $p_1 < \lambda p_0
+ (1 - \lambda)$ and the keyword does not occur.  The power curve includes the points
$(p_0, p_1)$ and $(1-p_0, 1-p_1)$.

The key point is that a mixture is not the unique way to implement the probability
distributions.  As we illustrate in Section~\ref{sec:class-prob-quant}, the superposed
vectors\begin{equation}
  \label{eq:25}
  \ket{\varphi_1} =
  \left(
    \begin{array}{c}
      \sqrt{p_1}\\ [5pt]
      \sqrt{1-p_1}
   \end{array}
  \right)
  \qquad
  \ket{\varphi_0} =
  \left(
    \begin{array}{c}
      \sqrt{p_0}\\ [5pt]
      \sqrt{1-p_0}
    \end{array}
  \right)
\end{equation}
 yield the pure densities
\begin{equation}
  \label{eq:2}
  \rho_1     = 
  \left(
    \begin{array}{cc}
      p_1 & \sqrt{p_1(1-p_1)}\\ [5pt]
      \sqrt{p_1(1-p_1)} & 1-p_1
    \end{array}
  \right)
  = \ketbra{\varphi_1}{\varphi_1}
\end{equation}
\begin{equation}
  \label{eq:4}
  \rho_0 =     
  \left(
    \begin{array}{cc}
      p_0 & \sqrt{p_0(1-p_0)}\\ [5pt]
      \sqrt{p_0(1-p_0)} & 1-p_0
    \end{array}
  \right)
  = \ketbra{\varphi_0}{\varphi_0}
\end{equation}
which replace the mixed densities.  Theorem~\ref{the:helstrom} instructs us to define the
optimal projectors as those of the spectrum of~\eqref{eq:5} whose eigenvalues are
positive, the spectrum being
\begin{equation}
  \eta_0 \mathbf{Q}_0 + \eta_1 \mathbf{Q}_1 \qquad
  \mathbf{Q}_0 = \ketbra{\eta_0}{\eta_0} \qquad
  \mathbf{Q}_1 = \ketbra{\eta_1}{\eta_1} 
\end{equation}
where the $\eta$'s are eigenvalues,
\begin{equation}
  \label{eq:36}
  \eta_0 = -R + \frac{1}{2}(1-\lambda) < 0
  \qquad
  \eta_1 = +R + \frac{1}{2}(1-\lambda) > 0
\end{equation}
and
\begin{eqnarray}
  R &=& \sqrt{\frac{1}{4}(1-\lambda)^+\lambda(1-\interm)} \\ \interm &=& \sqrt{p_0}\sqrt{p_1} +
\sqrt{1-p_0}\sqrt{1-p_1} \\ &=& |\braket{\varphi_0}{\varphi_1}|^2\label{eq:39}
\end{eqnarray}
(see~\cite{Helstrom76}).  $\interm$ is the distance between densities defined
in~\cite{Wootters81}; $\interm$ is the squared cosine of the angle between the subspaces
corresponding to the density vectors. The justification of viewing $\interm$ as a distance
comes from the fact that ``the angle in a Hilbert space is the only measure between
subspaces, up to a costant factor, which is invariant under all unitary transformations,
that is, under all possible time evolutions.''~\cite{Wootters81}

\begin{definition}
  $\mathbf{Q}_0, \mathbf{Q}_1$ are the optimal projectors in the pure case and
  $\mathbf{P}_0, \mathbf{P}_1$ are the optimal projectors in the mixed case.
\end{definition}

The probability of detection (i.e., the power) $Q_d$ and the probability of false alarm
(i.e., the size) $Q_0$ in the pure case are defined as follows:
\begin{eqnarray}
  \label{eq:9}
  Q_d &=& \frac{\eta_1 + \lambda(1-\interm)}{2R}
  \\
  Q_0 &=& \frac{\eta_1 - (1-\interm)}{2R}
\end{eqnarray}
Finally, $Q_d$ can be defined as function of $Q_0$:
\begin{equation}
  \label{eq:15}
  Q_d = 
  \left\{
    \begin{array}{ll}
      \left(\sqrt{Q_0}\sqrt{\interm} + \sqrt{1-Q_0}\sqrt{1-\interm}\right)^2 & 0 \leq
      Q_0 \leq \interm \\ [5pt]
      1 & \interm < Q_0 \leq 1
    \end{array}
  \right.
\end{equation}
so that the power curve is obtained (see ~\cite{Helstrom76}).  

Expressions~\eqref{eq:2} and~\eqref{eq:4} have no counterpart in classical probability and
are among the essential points of the paper because they allow us to improve ranking yet
using the same amount of evidence as the evidence used in the classical probability
distribution~\eqref{eq:6} and~\eqref{eq:7}.  

At this point, there are three main issues:
\begin{itemize}
\item the numerical difference between the classical and the quantum probabilities of
  detection at every given probability of false alarm,
\item the interpretations of $\mathbf{P}_0, \mathbf{P}_1$ and $\mathbf{Q}_0, \mathbf{Q}_1$
  and whether the interpretations can be tied together,
\item how the optimal projectors $\mathbf{Q}_0, \mathbf{Q}_1$ in the pure case can
  be used for ranking information units in a data management system.
\end{itemize}
The issues are addressed in
Section~\ref{sec:optim-rank-quant},~\ref{sec:interpr-quant-proj} and~\ref{sec:impl-rank},
respectively.



\section{Optimal Projectors in the Quantum Space}
\label{sec:optim-rank-quant}
\begin{figure}[t]
  \centering
  \includegraphics[width=0.750\columnwidth]{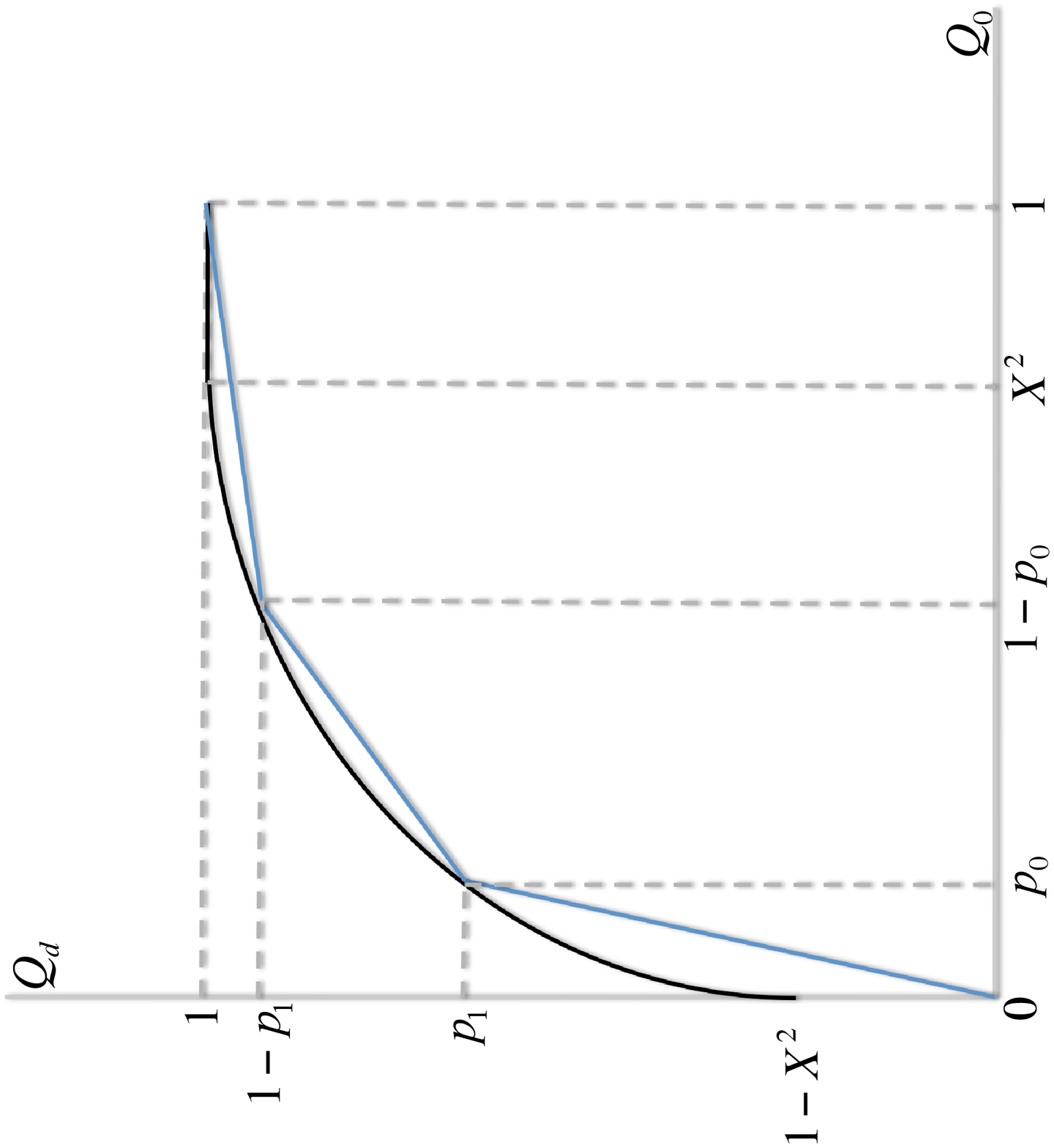}
  \caption{A graphical representation of the ROC curves. $Q_d$ is the curve above the
    polygonal curve depicting $P_d$.  The classical probability ROC curve intercepts the
    quantum probability ROC curve at $(p_0, p_1)$, $(1-p_0, 1-p_1)$ and $(1-p_0, 1-p_1)$,
    $(1, 1)$ where $\mathbf{P}_1, \mathbf{P}_0$ are observed. $Q_d = 1$ for $Q_0 >
    \interm$ and $Q_d \geq P_d$ for all $Q_0$'s.}
  \label{fig:roc}
\end{figure}

The following lemma shows that the power of the decision rule in quantum probability is
greater than, or equal to, the power of the decision rule in classical probability with
the same amount of information available from the training set to estimate $p_0, p_1$.
\begin{lemma}
  \label{sec:optim-rank-quant-2}
  $Q_d \geq P_d$ at every given false alarm probability.
\end{lemma}
\begin{proof}
  The equality holds only if $\mathbf{P}_i = \mathbf{Q}_i, i=0, 1$:
  \begin{equation*}
   \tr(\mathbf{\rho}_1\mathbf{P}_1) =
   \left(
     \begin{array}{cc}
       1 & 0
     \end{array}
   \right)
   \rho_1
   \left(
     \begin{array}{c}
       1 \\ [5pt] 0
     \end{array}
   \right)
   = 
   p_1
   = \tr(\mu_1\mathbf{P}_1)
  \end{equation*}
  \\
  \begin{equation*}
    \tr(\mathbf{\rho}_0\mathbf{P}_1)
    =  
    \left(
      \begin{array}{cc}
        1 & 0
      \end{array}
    \right)
    \rho_0
   \left(
      \begin{array}{c}
        1 \\ [5pt] 0
      \end{array}
    \right)
    = p_0
    = \tr(\mu_0\mathbf{P}_1)
  \end{equation*}
  Let $x \equiv P_0$ be a certain false alarm probability and let $Q_d(x), P_d(x)$ be the
  real, continuous functions yielding the detection probabilities at $x$.  $Q_d$ admits
  the first and the second derivatives in the range $[0,1]$. In particular, $Q_d'' < 0$ in
  $[0,1]$. $P_d$ is a continuous function. Consider the polynomial $L_0(x)$ of order $1$
  passing through the points $(0, 1-\interm)$ and $(p_0, p_1)$ at which $L_0$ intersects
  $Q_d$. Then, the Lagrange interpolation theorem can be used so that
  \begin{equation*}
    Q_d(x) - L_0(x) = Q''_d(c)x(x-p_0)/2
  \end{equation*}
  the latter being non negative because $Q_d'' < 0$ and $0 \leq x \leq p_0$. The number $c \in
  [0,p_0]$ exists due to the Rolle theorem. As $L_0(x) \geq P_d(x), x \in [0,p_0]$, hence,
  $Q_d(x) \geq P_d(x) , x \in [0,p_0]$.  Similarly, consider the polynomials $L_1(x)$ and
  $L_2(x)$ of order $1$ passing through the points $(p_0, p_1)$, $(1-p_0, 1-p_1)$ and $(1-p_0,
  1-p_1)$, $(1, 1)$ at which $L_1$ and $L_2$ intersect $Q_d$, respectively. Then, the Lagrange
  interpolation theorem can again be used so that
  \begin{equation*}
    Q_d(x) - L_1(x) = Q''_d(c)(x-p_0)(x-1+p_0)/2
  \end{equation*}
  \begin{equation*}
    Q_d(x) - L_2(x) = Q''_d(c)(x-1+p_0)(x-1)/2
  \end{equation*}
  Then, $Q_d(x) \geq P_d(x)$ for all $x \in [0,1]$.
\end{proof}
The power $Q_d$ can plotted against the size $Q_0$, thus producing the power curve of the
classical decision rule and the power curve of the quantum decision.  A graphical
representation is provided in Figure~\ref{fig:roc}.
\begin{example}
  \label{sec:optim-proj-quant}
  Suppose that ten information units have been used for training a data management system. Each
  unit has been indexed using one binary feature and has been marked as useful (1) or
  useless (0). The training set is summarized by Table~\ref{tab:example}:
  \begin{table}[!h]
    \centering
    \begin{equation*}
      \begin{array}{|l|c|c|c|c|c|c|c|c|c|c|}
        \hline
        unit	& u_1 & u_2 & u_3 & u_4 & u_5 & u_6 & u_7 & u_8 & u_9 & u_{10}\\
        \hline
        feature	& 1   & 1   & 1  & 1   & 1   & 0  & 0  & 0  & 0  & 0  \\
        use     & 1   & 1   & 1  & 0   & 0   & 1  & 0  & 0  & 0  & 0 \\
        \hline
      \end{array}      
    \end{equation*}
   \caption{The training set of Example~\ref{sec:optim-proj-quant}}
    \label{tab:example}
  \end{table}
  \\
  Therefore,
  \begin{equation*}
    p_1 = \frac{3}{5} \qquad p_0 =\frac{1}{5} \qquad \interm=0.91 \qquad d^2=0.09\ .
  \end{equation*}
  The computation of $P_d, P_0$ follows from~\eqref{eq:21} and the computation of $Q_d,
  Q_0$ follows from~\eqref{eq:9}.  When $P_0=Q_0$, we have that $Q_d=P_d$.
\end{example}

\section{Interpretation of the Optimal Projectors in Data Management}
\label{sec:interpr-quant-proj}

In this section, some interpretations of the optimal projectors representing the region of
acceptance are provided.  The optimal projectors $\mathbf{Q}_0, \mathbf{Q}_1$ in the pure
case have a more difficult interpretation than $\mathbf{P}_0, \mathbf{P}_1$ because the
latter represent ``physical'' observations (e.g., a customer review does exist or does
not) whereas $\mathbf{Q}_0, \mathbf{Q}_1$ cannot be expressed in terms of $\mathbf{P}_0,
\mathbf{P}_1$ and we cannot explain the $\mathbf{Q}_i$'s by saying that, for example, they
represent the presence and/or the absence of a feature.  In quantum theory, the
impossibility of expressing a projectors as functions of other projectors is termed
incompatibility which is expressed mathematically as $\mathbf{Q}_i \mathbf{P}_j \neq
\mathbf{P}_j \mathbf{Q}_i, \mathbf{P}_i \neq \mathbf{Q}_i$.

The interpretation of the optimal projectors reflects on the interpretation of what means
that they are ``observed'' in an information unit; for example, if the information unit is
a commercial item suggested to a customer, what does the ``observation'' of $\mathbf{Q}_1$
mean? What should we observe from an information unit so that the observation
outcome corresponds to the projector?  The question is not a futile because the answer(s)
would effect the algorithms (e.g., automatic indexing) used for representing the
informative content of the unit.  Specifically, the interpretation of $\mathbf{Q}_0,
\mathbf{Q}_1$ provides what the retrieval algorithm must do when a feature is observed.
Either the interpretation of an optimal projector is implemented at indexing time or at
query time, an internal memory representation in terms of data structures is necessary for
automatic processing and the representation needs the observation of physical properties
which are then converted into numbers.  In the mixed case, the answer is quite
straightforward because the optimal projectors correspond to the feature occurrence and
separate the units indexed by the feature from those not indexed.  In the pure case, the
answer is not straightforward at all. If $\mathbf{P}_0, \mathbf{P}_1$ represent feature
occurrence, the $\mathbf{Q}$'s cannot be a feature occurrence, but they are something new which
cannot be described in convential way.

\begin{figure}
  \centering
  \includegraphics[width=0.40\columnwidth]{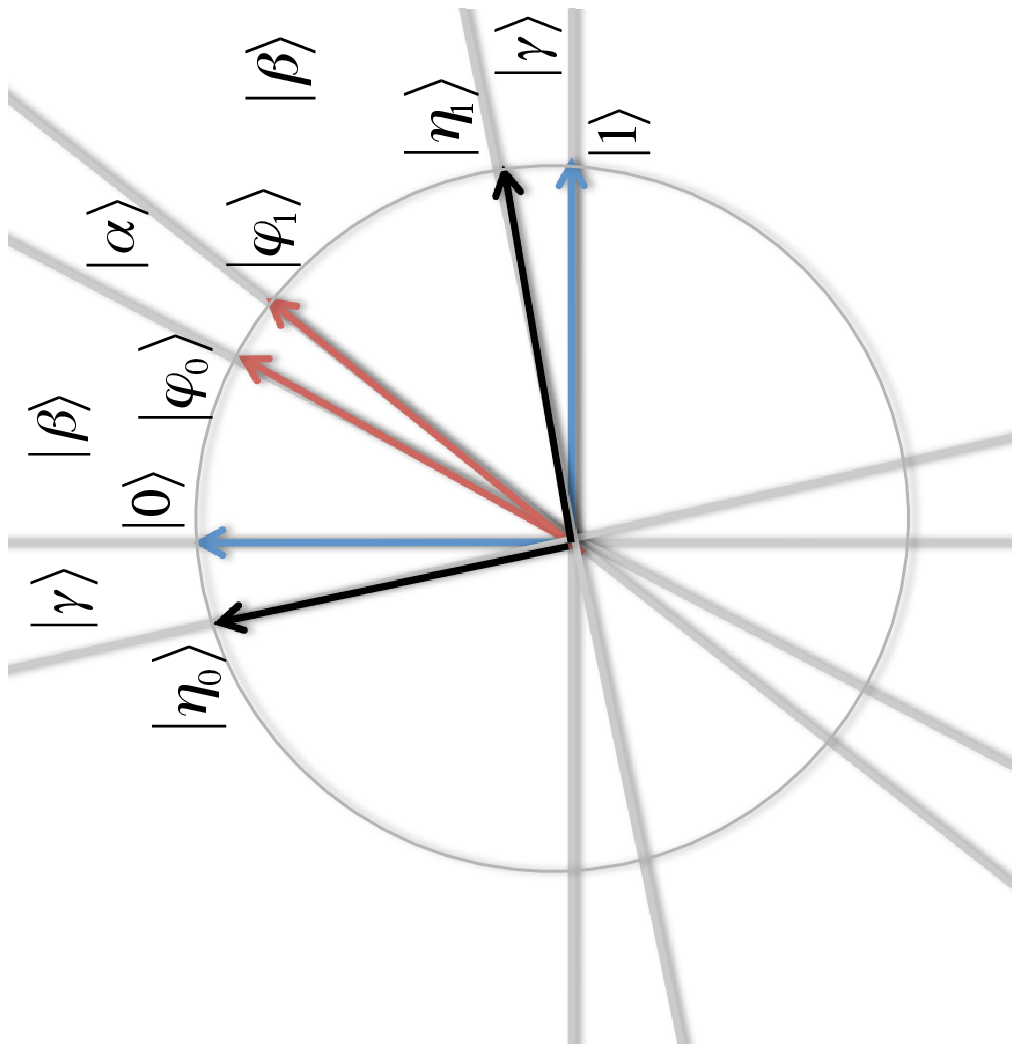}
  \caption{The geometric view of the canonical vectors in the mixed case, the optimal
    vectors in the pure case and the vectors representing the densities. The Greek letters
    $\alpha, \beta, \gamma$ indicate the angles between the vectors. }
  \label{fig:geometry}
\end{figure}
Geometrically, each vector is a superposition of other two independent vectors.
Figure~\ref{fig:geometry} depicts the way the vectors and the spanned subspaces (i.e.,
projectors) interact and shows that $\ket{\eta_1}, \ket{\eta_0}$ are placed symmetrically
``around'' the density vectors and the probability of error is
minimized~\cite{Helstrom76}.  The observation of a binary feature places the observer upon
either $\ket{0}$ or $\ket{1}$ and there is no way to move upon $\ket{\eta_0}$ or
$\ket{\eta_1}$.  

Probabilistically, the optimal projectors and the density vectors are related as follows:
\begin{eqnarray}
  \label{eq:29}
  \ket{\varphi_0} &=& \sqrt{1-Q_0}\ket{\eta_0}+\sqrt{Q_0}\ket{\eta_0}\\
  \ket{\varphi_1} &=& \sqrt{1-Q_d}\ket{\eta_0}+\sqrt{Q_d}\ket{\eta_1}
\end{eqnarray}
where
\begin{eqnarray}
  \label{eq:30}
  Q_d &=& \left|\braket{\eta_1}{0}\braket{0}{\varphi_1} + \braket{\eta_1}{1}\braket{1}{\varphi_1}\right|^2\\
  Q_0 &=& \left|\braket{\eta_1}{0}\braket{0}{\varphi_0} + \braket{\eta_1}{1}\braket{1}{\varphi_0}\right|^2
\end{eqnarray}

Logically, the projectors are assertions, thus a parallel can be established with
assertions and subsets~--~an assertion defines the elements of the universe (e.g., an
event space) which belong to a subset. The basic difference between subspaces and subsets
is that the vectors belong to a subspace if and only if they are spanned by a basis of the
subspace. A containment relationship can be established between subspaces such that  if a
subspace (e.g., a line) includes a point, then every subspace (e.g., a plane) containing
the line includes the point too.  The subspace spanned by the projector $\mathbf{A}$ is
termed as $L(\mathbf{A})$ and the containment relationship between $L(\mathbf{A})$ and
$L(\mathbf{B})$ can be encoded as $\mathbf{A}\mathbf{B} = \mathbf{B}$ such that  for every
vector $\ket{x}$, $\mathbf{A}\mathbf{B}\ket{x} =
\mathbf{B}\ket{x}$~\cite[Ch. 5]{vanRijsbergen04}.  

The paper considers the information units relevant, useful or interesting when they are
included by the subspace $L(\ket{\varphi_1})$ spanned by $\ket{\varphi_1}$.  Suppose that a
subspace $L(\mathbf{A})$ is given and that $|\braket{x-y}{x-y}|^2$ is the metric defined
on the whole space to which $L(\mathbf{A})$ belongs.
\begin{proposition}
  $\ket{y^*} = \arg_{\ket{y} \in L(\mathbf{A})} \min |\braket{x-y}{x-y}|^2 = \mathbf{A}\ket{x}$
\end{proposition}
\begin{proof}
  See~\cite{Melucci08a,vanRijsbergen04}.
\end{proof}
As $|\braket{x-y^*}{x-y^*}|^2 = \tr(\ketbra{x-y^*}{x-y^*}) = 1 - |\braket{x}{y^*}|^2$,
$\ket{y^*}$ maximizes the probability
\begin{equation}
  \tr(\ketbra{x}{y^*}) = |\braket{x}{y^*}|^2
  \label{eq:17}
\end{equation}
Note that  when $\ket{x}$ is in $L(\ket{\varphi_0})$,~\eqref{eq:17} is the distance
between the $L(\ket{\varphi_0})$'s defined in~\cite{Wootters81}.

The result establishes a connection between the geometric, probabilistic and logical
interpretation of the projectors even though they seems different. In the next section,
these interpretations are tied together, thus allowing us to look for a criterion for
assigning an information unit to the best region as explained.

\section{Implementation of the Optimal Ranking}
\label{sec:impl-rank}

The problem is to decide whether an information unit represented by a binary
feature\footnote{We recall that the binary case has been introduced for the sake of
  clarity.} is considered relevant, useful, interesting, etc..  An algorithm
implementing such a decision rule should perform as follows. It reads the feature
occurrence symbol (i.e., either $0$ or $1$); check whether the feature is included by the
region of acceptance. If the feature is not included, the hypothesis of relevance,
usefulness, interest, etc. is rejected.

Another view of the preceding decision rule is the ranking of the information
units. When ranking information units, the system returns the units whose features lead to
the highest probability of detection, then those whose features lead to the second highest
probability of detection, and so on.  When a binary feature is considered, the ranking
ends up to placing the units whose features lead to the highest probability of detection
whereas the other units are not retrieved.

As we point out in Section~\ref{sec:interpr-quant-proj}, the observation of features
corresponding to $\mathbf{P}_1, \mathbf{P}_0$ cannot give any information about the
observation of the events corresponding to $\mathbf{Q}_1, \mathbf{Q}_0$ due to the
incompatibility between these pairs of events.  Thus, we cannot design an
algorithm implementing the decision rule so that the observation of a feature can be
translated into the observation of the events corresponding to $\mathbf{Q}_0,
\mathbf{Q}_1$.

A possible approach can be based on the probabilistic interpretation of the optimal
projectors.  According to such an approach, the probability that the event corresponding to an
optimal projector occurs provided that a feature occurs can provide a measure of the degree to
which the event had occurred if it could have been observed.  When the subspaces represent
these events, the probability that the event corresponding to an optimal projector is observed
if the state is described by $\rho_1$ is $P_d$.  Such an approach is partially satisfactory
because $P_d \neq Q_d$.

As an alternative approach, consider the geometric interpretation depicted
in~Figure~\ref{fig:geometry}.  Note that the asymmetry of $\ket{0}, \ket{1}$ with respect to
the densities causes the suboptimality of $\mathbf{P}_1 = \ketbra{1}{1}, \mathbf{P}_0 =
\ketbra{0}{0}$.  Indeed, if $\ket{0}, \ket{1}$ coincided with $\ket{\eta_0}, \ket{\eta_1}$,
the power and the size would be the same in both cases.

We propose a method which reaches the optimality without neither resorting to
probabilistic approximations nor undergoing high computational costs.

The density vectors $\ket{\varphi_1}, \ket{\varphi_0}$ are a superposition of both
$\ket{0}, \ket{1}$ and $\ket{\eta_0}, \ket{\eta_1}$ which are two different bases and
induce different coordinates.  When $\ket{0}, \ket{1}$ is the basis, the coordinates of
$\ket{\varphi_i}$ are $\sqrt{p_i}, \sqrt{1-p_i}$.  When $\ket{\eta_0}, \ket{\eta_1}$ is
the basis, the coordinates of $\ket{\varphi_i}$ are $x_{00}, x_{01}, x_{10}, x_{11}$ such
that
\begin{equation}
  \label{eq:34}
  \ket{\varphi_0} = x_{00} \ket{\eta_0} + x_{01} \ket{\eta_1} 
  \qquad 
  \ket{\varphi_1} = x_{10} \ket{\eta_0} + x_{11} \ket{\eta_1} 
\end{equation}
and
\begin{eqnarray}
  \label{eq:35}
  x^2_{00} = \frac{\interm}{(1-\eta_1)^2+\interm} && x^2_{01} = \frac{(1-\eta_1)^2}{(1-\eta_1)^2+\interm}\\
  x^2_{10} = \frac{\interm}{(1+\eta_1)^2+\interm} && x^2_{11} = \frac{(1+\eta_1)^2}{(1+\eta_1)^2+\interm}
\end{eqnarray}
As $\lambda = 1$ is often assumed when ranking information units, the coordinates have a
quite simple and intuitive meaning provided by the following expressions:
\begin{eqnarray}
  \label{eq:37}
  x^2_{00} = \frac{1+d^2}{2} && x^2_{01} = \frac{1-d^2}{2} \label{eq:37a} \\
  x^2_{10} = \frac{1-d^2}{2} && x^2_{11} = \frac{1+d^2}{2} \label{eq:37b}
\end{eqnarray}
where $1 - d^2 = \interm$.

As the asymmetry of $\ket{0}, \ket{1}$ with respect the density vectors is due to $p_1,
p_0$ which summarize the statistics observed from the training set, we leverage them for
improving the ranking. In particular, in the paper we show that changing the estimation of
$p_1, p_0$ is sufficient to reach the optimality.

Then, we wonder how we should define the density vectors or matrices so that $Q_d, Q_0$
were obtained instead of $P_d, P_0$.  The basis vectors (i.e., $\ket{0}, \ket{1}$ or
$\ket{\eta_0}, \ket{\eta_1}$) are rotated, thus changing the coordinates.

Therefore, we define the density vectors $\ket{\varphi_0}, \ket{\varphi_0}$ in $\ket{0},
\ket{1}$ according to~\eqref{eq:34} in such a way that  if a feature is observed under
$H_1$, the probability of detection is $Q_d$ and, if a feature is observed under $H_0$,
the probability of false alarm is $Q_0$.  The simple solution is defining the new density
vectors as follows:
\begin{equation}
  \label{eq:38}
  \ket{\varphi_0'} = x_{00} \ket{0} + x_{01} \ket{1} 
  \qquad 
  \ket{\varphi_1'} = x_{10} \ket{0} + x_{11} \ket{1}   
\end{equation}
thus obtaining
\begin{equation}
  \label{eq:42}
  Q'_0 = \tr(\ketbra{\varphi_0'}{\varphi_0'}\mathbf{P}_1) = Q_0
  \qquad
  Q'_d = \tr(\ketbra{\varphi_1'}{\varphi_1'}\mathbf{P}_1) = Q_d
\end{equation}

At first sight, the increase of the probability of detection is due to the higher
probability values assigned to the region of acceptance in the pure case than those
assigned to the region of acceptance in the mixed case, and not to a different ranking. In
the following, we show that the superiority of the discriminant function in the pure case
is due to the different ranking induced by a different partition of the event space into
region of acceptance and region of rejection.

We state the problem as follows. Are there $p_0, p_1, \lambda$ such that the region of
acceptance in the pure case differs from that in the mixed case?  Consider
Theorem~\ref{the:helstrom} to answer the question.  The region of acceptance in the mixed
case is defined through Table~\ref{tab:region-of-acceptance-mixed} whereas the region of
acceptance in the pure case is defined through Table~\ref{tab:region-of-acceptance-pure}.
Furthermore, the discriminant function derived from~\eqref{eq:38} is
\begin{equation}
  \label{eq:41}
  \tr((\sigma_1 - \lambda\sigma_0)\mathbf{E}) \qquad \mathbf{E} \in \left\{ \mathbf{0},
      \mathbf{Q}_0, \mathbf{Q}_1, \mathbf{I} \right\}
\end{equation} 
where
\begin{equation}
  \label{eq:44}
  \sigma_i =
  \left(
    \begin{array}{cc}
      |x_{i1}|^2			& \sqrt{|x_{i1}|^2(1-|x_{i1}|^2)}	\\
      \sqrt{|x_{i1}|^2(1-|x_{i1}|^2)}	& 1-|x_{i1}|^2	
    \end{array}
  \right)
  \qquad i=0,1
\end{equation}

\begin{table}[t]
  \centering
  \begin{tabular}{|l|c|c|}
    \hline
    {}			& \multicolumn{2}{c|}{Second Eigenvalue}			\\
    \cline{2-3}
   {First Eigenvalue}	& $1-p_1<\lambda(1-p_0)$	& $1-p_1>\lambda(1-p_0)$	\\
    \hline
    $p_1<\lambda p_0$	& $\mathbf{0}$			& $\mathbf{P}_0$		\\
    $p_1>\lambda p_0$	& $\mathbf{P}_1$		& $\mathbf{I}$			\\
    \hline
  \end{tabular}
  \caption{The regions of acceptance corresponding to the sign of the eigenvalues of the
    spectrum of the discriminant function in the mixed case. The equality case is
    addressed in~\cite{Helstrom76}.}
  \label{tab:region-of-acceptance-mixed}
\end{table}

\begin{table*}[t]
  \centering
  \begin{tabular}{|l|c|c|}
    \hline
    {}			& \multicolumn{2}{c|}{Second Eigenvalue}	\\
    \cline{2-3}
    {First Eigenvalue}	& $1-|x_{11}|^2<\lambda(1-|x_{01}|^2)$	& $1-|x_{11}|^2>\lambda(1-|x_{01}|^2)$\\
   \hline
    $|x_{11}|^2<\lambda |x_{01}|^2$& $\mathbf{0}$		& $\mathbf{Q}_0$	\\
    $|x_{11}|^2>\lambda |x_{01}|^2$& $\mathbf{Q}_1$		& $\mathbf{I}$			\\
    \hline
  \end{tabular}
  \caption{The regions of acceptance corresponding to the sign of the eigenvalues of the
    spectrum of the discriminant function in the pure case. The equality case is
    addressed in~\cite{Helstrom76}.}
  \label{tab:region-of-acceptance-pure}
\end{table*}

Suppose that $p_0 = 1, p_1 = \frac{7}{10}, \lambda = \frac{1}{2}$. Thus, $X^2 =
\frac{7}{10}, R \cong 0.21, x_{11}^2 \cong 0.81, x_{01}^2 \cong 0.11$ and the region of
acceptance in the mixed case is represented by $\mathbf{I}$, whereas the region of
acceptance in the pure case is represented by $\mathbf{P}_1$.  The counter-example just
mentioned proves the following
\begin{corollary}
  \label{sec:impl-optim-rank}
  The discriminant function~\eqref{eq:41} ranks information units in a different way from
  the discriminant function~\eqref{eq:13} because an alternative ranking is computed.
\end{corollary}
In~\eqref{eq:13} the densities that are considered are those associated to a mixed state,
while~\eqref{eq:41} in the densities are the one associated to a pure state. So the equations
look like the same, what differs is the type of densities that are used in the two cases. We
have shown that the improvement of ranking measured in terms of probability of detection given
a probability of false alarm is due to the ranking induced by $Q'_0, Q'_d$.

Hence, we state the problem of finding $\mathbf{Q}_0, \mathbf{Q}_1$ into the problem of
defining the coordinates of the representation of the density vectors in $\ket{0},
\ket{1}$. The problem of defining the new coordinates for the density vectors might be
viewed as a problem of feature weighting.  In such a context, the traditional estimations
of the probability of feature occurrence (i.e., $p_1, p_0$) under two different hypothesis
are replaced by the $x_{ij}$'s.  Feature re-weighting is explored in IR whose
state-of-the-art is given by the BM25 weighting scheme surveyed, for example,
in~\cite{Robertson&09}.  The main drawback of the weighting schemes like BM25 is the
parameter tuning necessary for optimizing the effectiveness with a given database or
query, thus making the understanding of how and why a scheme is more effective than others
rather problematic.

In contrast, the paper illustrate the decision rule in such a way that if the decision rule is
followed, then $H_1$ shall be accepted when it is true at a higher probability of detection
than $P_d$ when the probability of false alarm is not more than a given threshold.  The
formulation of the decision rule provided in this section allows us to design an efficient
algorithm for indexing and retrieving (or classifying) information units.  The algorithm is
just an instance of those employed currently in IR (see~\cite{Croft&09} for example) where the
maximum likelihood or Bayesian estimations of $p_0, p_1$ are replaced by~\eqref{eq:37a}
and~\eqref{eq:37b}.


\section{Related Work}
\label{sec:related-work}

The foundations of quantum mechanics and theory has been illustrated in plenty of books such
as~\cite{Griffiths02} and~\cite{Hughes89}.  Quantum probability, for example, has been
introduced in~\cite{Parthasarathy92}.  In particular, the interference term is addressed
in~\cite{Accardi84}.  The view of probability illustrated in
Section~\ref{sec:class-prob-quant} is based on~\cite{Rieffel07}.  The utilization of quantum
theory in computation, information processing and communication is described
in~\cite{Nielsen&00}.  Recently, investigations have started in other research areas, for
example, in IR~\cite{Bruza&09}.

The paper is inspired by Helstrom's book~\cite{Helstrom76} which provides the foundations and
the main results in quantum detection; an example of the exploitation of the results in
quantum detection is reported~\cite{Cariolaro&10} within communication theory.  This paper
links to~\cite{vanRijsbergen04} as far it concerns density matrices and projectors; however,
the paper develops quantum detection for data management.

This paper departs from the Probability Ranking Principle (PR) proposed in the context of
classical probability; we propose quantum probability to improve ranking in a principled
way. In Information Retrieval, the Probability Ranking Principle (PRP) states that ``If a
reference retrieval system's response to each request is a ranking of the documents in the
collection in order of decreasing probability of relevance to the user who submitted the
request, where the probabilities are estimated as accurately as possible on the basis of
whatever data have been made available to the system for this purpose, the overall
effectiveness of the system to its user will be the best that is obtainable on the basis of
those data.''~\cite{Robertson77}.  However, some assumptions undermine the general
applicability of the PRP.  We state a similar principle yet replace classical probability,
which is implied in~\cite{Robertson77}, with quantum probability~--~parameter estimation data
are kept the same, bu we instead use subspaces to define alternative regions of acceptance and
rejection.

To our knowledge, the use of quantum probability for ranking information units has not yet
been addressed in the same way of this paper although a few papers that are somehow comparable
can be found. Perhaps, the closest paper is~\cite{Zuccon&09}.  That paper proposes to rank
documents by quantum probability and suggests that interference (which must be estimated)
might model dependencies in relevance judgements such that documents ranked until position $n
− 1$ interfere with the degree of relevance of the document ranked at position $n$.  This
means, the optimal order of documents under the PRP differs from that of the Quantum PRP. Note
that they empirically show that quantum probability is more effective that classical one in
specific rankings tasks. 

In contrast, in this paper, we do not need to address interference because quantum probability
can be estimated using the same data used to estimate classical probability.  We rather show
that not only ranking by quantum probability provides a different optimal ranking, it is also
more effective than classical probability.  With this regard, the effectiveness of quantum
probability measured in~\cite{Zuccon&09} stems from the estimation of classical probability
and that of interference.  But, the regions of acceptance and rejection are still based on
sets. It follows that the optimality of the Quantum PRP strongly depends on the optimality of
the PRP and on the interference estimantion effectiveness. In this paper, on the contrary,
ranking optimality only depends on the region of acceptance defined upon subspaces.

Another paper somewhat related to ours is~\cite{Piwowarski&10}.  The authors discuss how to
emply quantum formalisms for encompassing various Information Retrieval tasks within a single
framework.  From an experimental point of view, what that paper demonstrates is that ranking
functions based on quantum formalism are computationally feasible.  The best experimental
results of rankings driven by quantum formalism are comparable to BM25, that is, to PRP, thus
limiting the contribution within a classical probability framework.

Probabilistic databases systems manage imprecise data and provide tools for structured complex
queries.  A survey is provided in~\cite{Dalvi&09}.  Beside scalability and query plan
execution, these systems do probabilistic inference which may be defined upon classical or
quantum probability and they concentrate on top-\emph{k} query answering where the tuples are
assigned a probability distribution.  The results of this paper may be applied to
probabilistic databases systems too.


\section{Future Developments and Conclusions}
\label{sec:conclusions}

The main result of the paper is the demonstration that quantum probability can be incorporated
into a data management system for ranking information units. As ranking by quantum probability
is more effective than ranking by classical probability when it has been used in other
domains, it is our belief that an analogous improvement can be achieved within data
management.
 
The future developments are threefold.  First, we will work on the intepretation of the
optimal projectors in the pure case because the detection of them in an information unit may
open further insights.  Second, feature classical correlation and quantum entanglement will be
investigated.  Third, evaluation is crucial to understand whether the results of the paper can
be confirmed by the experiments.


\appendix

\section{Dirac Notation}
\label{sec:dirac-notation}

A complex vector $x$ is represented as $\ket{x}$ and is called ``ket''. The conjugate
transpose of $x$ is represented as $\bra{x}$ and is called ``bra'' (therefore, the Dirac
notation is called the bra(c)ket notation). 

The inner product between $x$ and $y$ is represented as $\braket{x}{y}$, which is a
complex number.  The inner procuct between $x$ and itself is $\braket{x}{x}$. 

The outer product (or dyad) is $\ketbra{x}{y}$. A special case of dyad is $\ketbra{x}{x}$
which is the projector made onto $x$. 

If $A$ is a matrix (or an operator), then $A\ket{x}$ is the vector resulting from the
linear transformation represented by $A$.  $\ketbra{x}{x}$ is also an operator because it
is a projector. 

The real number $|\braket{x}{y}|^2$ is the squared inner product between $x$ and $y$.

Moreover, $\bra{x}\mathbf{A}\ket{y} = \tr(\bra{x}\mathbf{A}\ket{y})$ and the properties of
trace allow us to write $\tr(\bra{x}\mathbf{A}\ket{y}) = \tr(\mathbf{A}\braket{x}{y})$.
In general, if $\mathbf{A}, \mathbf{B}$ are trace-1 Hermitian operators, and $\mathbf{B}$
is a projector, $0 \leq \tr(\mathbf{A}\mathbf{B}) \leq 1$ is the probability that the
event represented by $\mathbf{B}$ occurs given a density operator $\mathbf{A}$.

The Dirac notation allow us to write complex expressions in an elegant way, for example,
\begin{eqnarray*}
  \label{eq:40}
  \braket{x}{y}\braket{y}{z}\braket{z}{x} &=& \tr(\braket{x}{y}\braket{y}{z}\braket{z}{x})\\
  {} &=& \tr(\ketbra{x}{x}\ketbra {y}{y}\ketbra{z}{z}))\\
  {} &=& \tr(\braket{z}{x}\braket{x}{y}\braket{y}{z})
\end{eqnarray*}

When $\mathbf{A}$ is mixed and $\mathbf{E}$ is a projector,
\begin{eqnarray*}
  \tr(\mathbf{A}\mathbf{E}) &=& \tr\left(\left(\sum_{i=0}^{r-1} \alpha_i
    \mathbf{E}_i\right)\mathbf{E} \right) \\
  {} &=& \tr\left(\sum_{i=0}^{r-1} \alpha_i \mathbf{E}_i \mathbf{E} \right)\\
  {} &=& \sum_{i=0}^{r-1} \alpha_i \tr(\mathbf{E}_i \mathbf{E})
\end{eqnarray*}


\end{document}